\documentclass[prd,twocolumn]{revtex4}

\usepackage{graphicx}
\usepackage{dcolumn}
\usepackage{amsmath}
\usepackage {epsfig}
\usepackage{bm}
\usepackage{amssymb}
\usepackage{latexsym}

\def\setC{\mathbb{C}}

\def\setR{\mathbb{R}}

\newcommand{\lta}{\lesssim}
\newcommand{\gta}{\gtrsim}

\newcommand{\be}{\begin{equation}}
\newcommand{\en}{\end{equation}}
\newcommand{\bea}{\begin{eqnarray}}
\newcommand{\ena}{\end{eqnarray}}

\newcommand{\nS}{n_{_{\mathrm{S}}}}
\newcommand{\mP}{m_{_{\mathrm{Pl}}}}

\newcommand{\zero}{{_0}}

\newcommand{\ess}{{_\mathrm{S}}}
\newcommand{\ti}{{_\mathrm{T}}}
\newcommand{\muS}{\mu_\ess}
\newcommand{\muT}{\mu_\ti}
\newcommand{\muST}{\mu_{\ess,\ti}}
\newcommand{\omegaT}{\omega_\ti}
\newcommand{\omegaS}{\omega_\ess}
\newcommand{\omegaST}{\omega_{\ess, \ti}}

\newcommand{\calH}{\mathcal{H}}

\newcommand{\GReCO}{${\cal G}\setR\varepsilon\setC{\cal O}$}

\makeatother
\begin{document}

\title[Short Title]{Inflation and Precision Cosmology}
\author{J\'er\^ome Martin$^1$} 
\affiliation{$^1$ Institut d'Astrophysique de Paris, \GReCO--FRE 2435, 
98bis boulevard Arago, 75014 Paris, France} 

\date{\today}

\begin{abstract}
A brief review of inflation is presented. After having demonstrated
the generality of the inflationary mechanism, the emphasize is put on
its simplest realization, namely the single field slow-roll
inflationary scenario. Then, it is shown how, concretely, one can
calculate the predictions of a given model of inflation. Finally, a
short overview of the most popular models is given and the
implications of the recently released WMAP data are briefly (and
partially) discussed.

\end{abstract}


\maketitle

\section{Introduction}

The inflationary scenario~\cite{inflation} has been invented in order
to solve and explain some observational facts (isotropy of the Cosmic
Microwave Background Radiation--CMBR--, flatness of the space-like
sections, etc \dots ) that could not be properly understood in the
context of the standard hot Big Bang theory. Therefore, at the
beginning of its history, the inflationary scenario was only able to
make postdictions but, given the problems of the standard model at
that time, this was already a success. However, soon after its advent,
it was realized~\cite{MuChi,pert} that inflation, when combined with
quantum mechanics, can also give a very convincing mechanism for
structure formation. In particular, for the first time, it was
understood how to generate a scale-invariant power
spectrum. Therefore, the inflationary mechanism was able to establish
a beautiful connection between facts which, before, were considered as
independent. However, since the Harisson-Zeldovich was already known
to be in agreement with the observations, it could be argued that,
somehow, this was again a postdiction. In fact, the inflationary
scenario does not predict a scale invariant spectrum but a nearly
scale invariant spectrum, the deviations from the scale invariance
being linked to the microphysics description of the theory. This
constitutes a definite prediction of inflation that can be
tested~\cite{LLMS}.

\par

Slightly more than twenty years after the invention of inflation, the
situation has recently changed because, thanks to the high accuracy
CMBR data obtained, among others, by the WMAP satellite~\cite{wmap},
we can now start to probe the details of the inflationary scenario and
check its predictions. The goal of this short review is, after having
tried to justify why the inflationary mechanism is generic (section
I), to show how, in its most popular and simplest realization, one can
calculate concrete predictions (section II) and compare them with the
recent observations (section III). These proceedings, due to the lack
of space, do not cover many important topics. Among them and since
this is particularly relevant for the present article, we just would
like to signal the discussion of why the presence of the CMBR Doppler
peaks strongly suggests that a phase of inflation took place in the
early universe, see Ref.~\cite{dod}.

\section{The inflationary mechanism}

\subsection{Basic equations}

The cosmological principle implies that the universe is, on large
scales, homogeneous and isotropic. This simple assumption drastically
constrains the possible shapes of the Universe which can solely be
described by the Friedman-Lema\^{\i}tre-Robertson-Walker (FLRW)
metric: ${\rm d}s^2=-{\rm d}t^2+a^2(t)\gamma _{ij}^{(3)} {\rm
d}x^i{\rm d}x^j$,
where $\gamma _{ij}^{(3)}$ is the metric of the three-dimensional
space-like sections of constant curvature. The time-dependent function
$a(t)$ is the scale factor. If the spatial curvature vanishes then
$\gamma _{ij}^{(3)}=\delta _{ij}$, where $\delta _{ij}$ is the
Kr\"onecker symbol. The previous expression is written in terms of the
cosmic time $t$ but it is also interesting to work in terms of the
conformal time $\eta $ defined by ${\rm d}t=a(\eta ){\rm d}\eta
$. Then, the metric can be re-expressed as
\begin{equation}
{\rm d}s^2=a^2(\eta )\left[-{\rm d}\eta ^2+
\gamma ^{(3)}_{ij}{\rm d}x^i{\rm d}x^j\right]\, .
\end{equation}
In terms of conformal time, the Hubble parameter $H=\dot{a}/a$ can be
written as $H={\cal H}/a$ where ${\cal H}\equiv a'/a$, a dot denoting
a derivative with respect to cosmic time while a prime stands for a
derivative with respect to conformal time.

\par

The matter is assumed to be a collection of $N$ perfect fluids
and, as a consequence, its stress-energy tensor is given by the
following expression
\begin{equation}
T_{\mu \nu }=\sum _{i=1}^NT_{\mu \nu}^{(i)}=(\rho +p)u_{\mu
}u_{\nu}+pg_{\mu \nu}\, ,
\end{equation}
where $\rho $ is the (total) energy density and $p$ the (total)
pressure. These two quantities are linked by the equation of state,
$p=\omega (\rho )$ [in general, there is an equation of state per
fluid considered, {\it i.e.} $p_i=\omega _i(\rho _i)$]. The vector
$u_{\mu }$ is the four velocity common to all fluids and satisfies the
relation $u_{\mu }u^{\mu }=-1$. In terms of cosmic time this means
that $u^{\mu }=(1,0)$ whereas in terms of conformal time one has
$u^{\mu }=(1/a,0)$ and $u_{\mu }=(-a,0)$. The fact that the
stress-energy tensor is conserved, $\nabla ^{\alpha }T_{\alpha
\mu}=0$, implies $\rho '+3{\cal H}(\rho +p)=0$.
This expression is obtained from the time-time component of the
conservation equation. The time-space and space-space components do
not lead to any interesting equations for the background.

\par

We are now in a position to write down the Einstein equations which
are just differential equations determining the scale factor. In terms
of cosmic time, they read
\begin{equation}
\label{Ecosmic}
\frac{\dot{a}^2}{a^2}+\frac{k}{a^2} =
\frac{\kappa }{3}\rho \, , 
\quad
-\biggl(2\frac{\ddot{a}}{a}+\frac{\dot{a}^2}{a^2}+\frac{k}{a^2}\biggr) 
=\kappa p \, ,
\end{equation}
where $\kappa \equiv 8\pi G=8\pi /\mP^2$, $\mP$ being the Planck mass
and where $k=0,\pm 1$ is the normalized three-dimensional
curvature. Combining the two expressions above, one obtains an
equation which permits to express the acceleration of the scale factor
$\ddot{a}/a=\kappa (\rho +3p)/6$.
From the above equation, one sees that any form of matter such that
$\rho +3p<0$ will cause an acceleration of the scale factor (this is
only true, of course, if the matter component satisfying $\rho +3p<0$
is the dominant one). The energy density is always positive but, in
some situation, the pressure can be negative and the inequality $\rho
+3p<0$ may be realized. This simple remark is at the heart of the
inflationary scenario. Let us notice that the above property is deeply
rooted into the fundamental principles of general relativity. It is
because all forms of energy weighs in general relativity that the
pressure participates to the equation giving the expression of
$\ddot{a}$. This has to be contrasted with Newtonian physics where the
same expression reads $\ddot{a}/a=-\kappa \rho/6 $, {\it i.e.} only
the energy density affects the expansion. As a consequence, the
expansion can only be decelerated. 

\par

A class of solutions of particular interest is that for which the
equation of state $\omega \equiv p/\rho $ is a constant. In this case,
the conservation equation can be immediately integrated and leads to
$\rho \propto a^{-3(1+\omega )}$.
Then, the scale factor is given by a power law either of the conformal
or of the cosmic time, namely
\begin{equation}
a(\eta )=\ell _\zero\vert \eta \vert ^{1+\beta }, \quad 
a(t)=a_\zero\biggl\vert \frac{t}{t_\zero }\biggr \vert^p\, ,
\end{equation}
where the sign of the conformal and cosmic times, to be discussed
below, can be negative or positive. The parameters $p$ and $\beta $
are related by $\beta =(2p-1)/(1-p)$ and $p=(1+\beta )/(2+\beta )$.
The link between $\omega $ and the parameters $\beta $ and $p$ can be
expressed as
\begin{equation}
\omega =\frac{1-\beta }{3(1+\beta )}=-1+\frac{2}{3p}\, .
\end{equation}
The cases $\beta =-1$ and/or $p=0$ are obviously meaningless since
they do not correspond to a dynamical scale factor.

\par

Let us now come back to the question of the sign of the conformal and
cosmic times. For the class of models under consideration, the Hubble
parameter can be easily calculated and is given by $H=p/t$. Let us
first consider the case where we have an expansion, {\it i.e.}
$H>0$. In this case, one has $t>0$ for $p>0$ and $t<0$ for $p<0$. On
the other hand, in the case of a contraction, $H<0$, we have $t>0$ for
$p<0$ and $t<0$ for $p>0$.  The link between the two times, ${\rm
d}t=a{\rm d}\eta $, takes the form
\begin{eqnarray}
\eta &=& \frac{1}{a_\zero (1-p)}\vert t\vert ^{1-p}, \quad \mbox{if} 
\quad t>0\, ,
\\
\eta &=& \frac{-1}{a_\zero (1-p)}\vert t\vert ^{1-p}, \quad \mbox{if} 
\quad t<0\, ,
\end{eqnarray}
from which we deduce that, if $t>0$, then $\eta <0$ for $p>1$ and
$\eta >0$ for $p<1$ and that, if $t<0$, then $\eta >0$ for $p>1$ and
$\eta <0$ for $p<1$.

\par

Finally, let us also examine some cases of particular interest. The
case $p=2/3$ corresponds to an expanding matter dominated universe and
$\omega =0$. Therefore, we have $\eta >0$ and $t>0$. The same
conclusion applies for the case $p=1/2$, {\it i.e.} the case of an
expanding radiation dominated universe with $\omega =1/3$. The case
$\beta =-2$ corresponds to $p=\infty $, that is to say to an
exponential scale factor: this is just the de Sitter solution with
$\omega =-1$. For $\beta <-2$, one has $p>1$ and $\eta <0$, $t>0$ if
we are interested in the case of expansion.

\par

The evolution of the background space-time can be very roughly
understood by means of the previous set of simple equations. The hot
Big Bang model before the advent of inflation just consisted into a
radiation dominated epoch ($\omega =1/3$) taking place at high
redshifts, until $z_{\rm eq }\simeq 10^4$, followed by a phase
dominated by cold matter ($\omega =0$). However, already at this
level, this very simple framework leads to unacceptable
conclusions. We now describe what are (some of) the problems of the
pre-inflationary hot Big Bang scenario and how postulating a new phase
of accelerated expansion in the very early universe can avoid these
problems.

\subsection{The horizon problem}

The horizon problem consists in the following. The furthest event that
we can directly ``see'' in the universe is the (re)combination, {\it
i.e.} the time at which the electrons and the protons combined to form
hydrogen atoms. Since the photon--atom cross-section (Rayleigh
cross-section) is much smaller than the photon--electron cross-section
(Thomson cross-section), the universe became transparent at that
time. The COBE~\cite{cobe} and WMAP~\cite{wmap} maps of the sky are
photographs of the universe at this epoch. The recombination took
place at a redshift of $z_{\rm lss}\simeq 1100$, {\it i.e.} within the
epoch dominated by the cold matter. Before, the universe was opaque
and therefore it is not possible to observe it directly at earlier
times from the Earth. Since no physical process can act on scales
larger than the horizon (see below for a precise definition of the
horizon), we typically expect the universe to be strongly
inhomogeneous on those scales. Seen from the Earth, this means that
the COBE map should look extremely different on angular scales larger
than the angular scale of the horizon at recombination.

\par

In order to investigate the consequences of the above statement, let
us calculate the angular diameter of the horizon at recombination,
seen by an observer today. Roughly speaking, this is just the size of
the horizon at the last scattering surface divided by the present
(angular) distance to the last scattering surface. In other words, it
can be expressed as $\Delta \Omega = d_{_{\rm H}}(t_{\rm lss})
/d_{_{\rm A}}(t_{\rm lss})$,
where we now discuss precisely the meaning of the terms in the above
formula.

\par

For this purpose, it is convenient to choose the coordinates system
such that the origin is located on Earth, {\it i.e.} such that ``our''
co-moving coordinate is $r=0$.  Suppose that a photon is emitted at
spatial co-moving coordinates $(r_{\rm em},\theta _{\rm em}, \varphi
_{\rm em})$ and at cosmic time $t_{\rm em}$. The path followed by the
photon can be chosen such that $\theta =\mbox{cst}. $ and $\varphi
=\mbox{cst}$ since this is a solution of the geodesic equation. In
this case, the path is completely characterized by the function
$r=r(t)$. This quantity is given by
\begin{equation}
\label{hor}
r(t)=r_{\rm em}-\int _{t_{\rm em}}^{t}\frac{{\rm d}\tau}{a(\tau)}
\Rightarrow 
d_{_{\rm P}}(t)=a(t)\biggl[
r_{\rm em}-\int _{t_{\rm em}}^{t}\frac{{\rm d}\tau}{a(\tau)}
\biggr]\, ,
\end{equation}
where $d_{_{\rm P}}(t)$ is the physical (proper) distance from the
``position'' of the photon at time $t$ to the origin. 

\par

This equation can be used to define the horizon. Indeed, the question
that one may ask is the following. At a given (reception) time,
$t=t_{\rm rec}$, what is the proper distance to the furthest point
where a photon, sent to us from there, could have reached the Earth
(the point of co-moving coordinate $r=0$) before or at the time $t_{\rm
rec}$? This proper distance is called the size of the horizon at time
$t=t_{\rm rec}$. Clearly the distance is maximized if the time of
emission is the Big-Bang and if the photon has just reached the Earth
at the time $t_{\rm rec}$. Hence the co-moving coordinate of emission
is obtained by writing that $d_{_{\rm P}}(t_{\rm rec})=0$ and by
taking a vanishing lower bound in the previous integral. This implies
that
\begin{equation}
r_{\rm em}=\int _{0}^{t_{\rm rec}}\frac{{\rm d}\tau}{a(\tau)}\, .
\end{equation}
This means that the distance to the horizon, at the time $t=t_{\rm
rec}$, is given by $d_{_{\rm H}}(t_{\rm rec}) =a(t_{\rm rec})r_{\rm
em}$.
In this equation, $t_{\rm rec}$ can be for instance the time of
recombination, $t_{\rm lss}$ or the present time $t_0$ depending on
whether one wants the evaluate the size of horizon at the last
scattering surface or now.

\par

Another question is to calculate the distance to a point where a
photon emitted at $t=t_{\rm em}$ has just arrived on Earth now, at
time $t=t_0$. Writing again that the photon is received now, one
obtains the corresponding co-moving coordinate of emission $r_{\rm
em}=\int _{t_{\rm em}}^{t_0} {\rm d}\tau/a(\tau)$.
From the previous equation, one can deduce that the corresponding
angular distance to the point of emission is given by $d_{_{\rm
A}}\equiv a(t_{\rm em})r_{\rm em}$. Indeed, the FLRW metric can be
written as (for flat space-like section) $ {\rm d}s^2=-c^2{\rm
d}t^2+a^2(t)({\rm d}r^2+r^2{\rm d}\Omega _2^2)$ and therefore the
proper distance $D$ across a source is $D\simeq ar\Delta \Omega $ at
time $t_{\rm em}$ (obtained from ${\rm d}t={\rm d}r=0$ since it is
supposed that the source is located on a sphere of radius
$r=\mbox{cte}$). As a consequence, one has $\Delta \Omega =D/(ar)$
from which we deduce $d_{_{\rm A}}\equiv a(t_{\rm em})r(t_{\rm
em})$. Notice that the proper distance to the point of emission is
$a(t_0)r_{\rm em}$.  For very high redshifts, as for instance $z_{\rm
lss}$, these two distances are of course very different.

\par

We can now deduce the general expression of the angular diameter. It
is given by
\begin{equation}
\displaystyle
\Delta \Omega =\biggl[\int _0^{t_{\rm lss}}
\frac{{\rm d}\tau}{a(\tau)}\biggr]\times \biggl[\int _{t_{\rm lss}}^{t_0}
\frac{{\rm d}\tau}{a(\tau)}\biggr]^{-1}\, .
\end{equation}
In the previous expression, the factors $a(t_{\rm lss})$ have canceled
out.  

\par

Let us now try to evaluate the above solid angle in a realistic case
where matter and radiation are present~\cite{Ellis}. For simplicity,
we assume that the universe is radiation dominated before
recombination and matter dominated after. In reality, as already
mentioned, equivalence between radiation and matter takes place before
the recombination but this does not introduce important
corrections. Since we are going to study the influence of a phase of
inflation, we also assume that the epoch dominated by radiation can be
interrupted during the period $t_{\rm i}<t<t_{\rm end}$. During this
interval, we assume that the universe is dominated by an unknown fluid
the equation of state of which is constant and given by $\omega
_{_{\rm X}}$. To recover the standard hot Big Bang case, where this
epoch does not occur, it is sufficient to consider that $t_{\rm
i}=t_{\rm end}$, {\it i.e.} to switch off the phase dominated by the
unknown fluid. The scale factor is not known exactly but its piecewise
expression reads:
\begin{widetext}
\begin{eqnarray}
\label{scalefactor}
a(t) &=& a_{\rm i}(2H_{\rm i}t)^{1/2}\, , 
\quad 0\le t<t_{\rm i}\, , 
\quad
a(t) = a_{\rm i}\biggl[\frac32(1+\omega _{_{\rm X}})
H_{\rm i}(t-t_{\rm i})+1\biggr]^{2/[3(1+\omega _{_{\rm X}})]}\, , 
\quad t_{\rm i}\le t<t_{\rm end}\, , 
\\
a(t) &=& a_{\rm end}[2H_{\rm end}(t-t_{\rm end})+1]^{1/2}\, , 
\quad t_{\rm end}\le t<t_{\rm eq}\, , 
\quad 
a(t) = a_{\rm eq}\biggl[\frac{3H_{\rm eq}}{2}
(t-t_{\rm eq})+1\biggr]^{2/3}, \quad t_{\rm eq}\le t<t_0 \, .
\end{eqnarray}
At each transitions, the scale factor and its first time derivative
are continuous. A straightforward calculation leads to the expressions
of the horizon at decoupling and of the angular distance to the last
scattering surface. One finds
\begin{eqnarray}
\label{angularlss}
d_{_{\rm A}}(t_{\rm lss}) &=& a_{\rm lss}\int _{t_{\rm lss}}^{t_0} 
\frac{{\rm d}\tau}{a(\tau)} 
=a_{\rm lss}\times \frac{2}{a_0H_0}\left[1-
\biggl(\frac{a_{\rm lss}}{a_0}\biggr)^{1/2}\right] \, ,
\\
\label{horizonlss}
d_{_{\rm H}}(t_{\rm lss}) &=& a_{\rm lss}
\int _{0}^{t_{\rm lss}} \frac{{\rm d}\tau}{a(\tau)} =
a_{\rm lss}\times 
\frac{1}{a_0H_0}\biggl(\frac{a_{\rm lss}}{a_0}\biggr)^{1/2} 
\left\{1+\frac{1-3\omega _{_{\rm X}}}{1+3\omega _{_{\rm X}}}
\frac{a_{\rm end}}{a_{\rm lss}}\left[1-
\biggl(\frac{a_{\rm i}}{a_{\rm end}}
\biggr)^{(1+3\omega _{_{\rm X}})/2}
\right]\right\}\, .
\end{eqnarray}
From the aboves equations, we deduce the expression of the solid angle
\begin{equation}
\label{solid}
\Delta \Omega =\frac12
\biggl[1-(1+z_{\rm lss})^{-1/2}\biggr]^{-1}(1+z_{\rm lss})^{-1/2}
\biggl \{1+\frac{1-3\omega _{_{\rm X}}}{1+3\omega _{_{\rm X}}}
\frac{1+z_{\rm lss}}{1+z_{\rm end}}\biggl[1-{\rm e}^{-N
(1+3\omega _{_{\rm X}})/2}\biggl]\biggr\}\, ,
\end{equation}
\end{widetext}
where $N\equiv \ln (a_{\rm end}/a_{\rm i})$ is the number of ${\rm
e}$-foldings during inflation and $z_{\rm end}$ is the redshift at
which inflation stops (corresponding to $t=t_{\rm end}$).

\par

Let us first suppose that there is no phase of inflation, {\it i.e.}
$N=0$. Then, $\Delta \Omega \simeq 0.5\times (1+z_{\rm lss})^{-1/2}
\simeq 0.85^{\circ}$.
\begin{figure*}
\begin{center}
\includegraphics[width=8.5cm, height=6.4cm, angle=0]{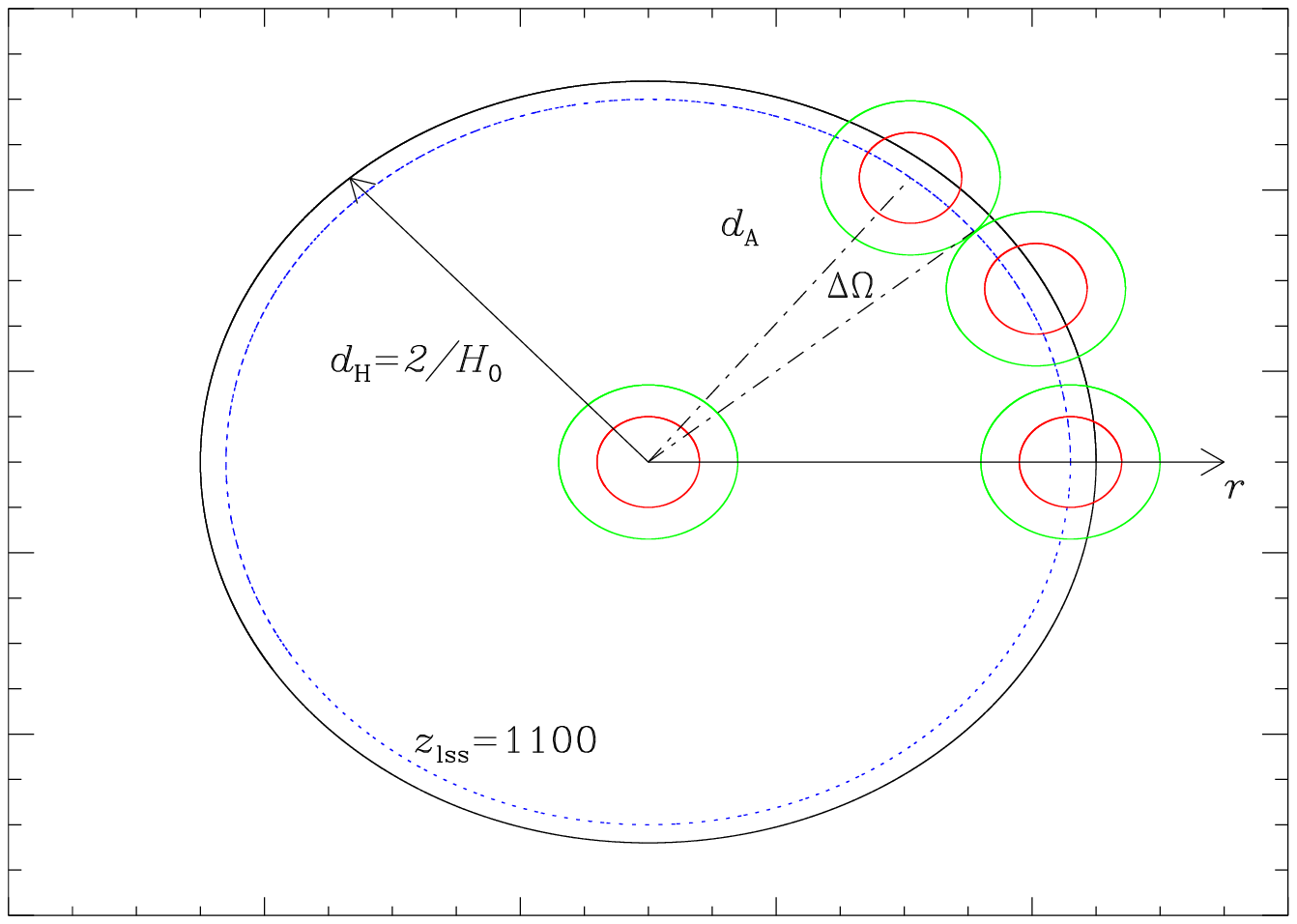}
\includegraphics[width=9.25cm, height=6.8cm, angle=0]{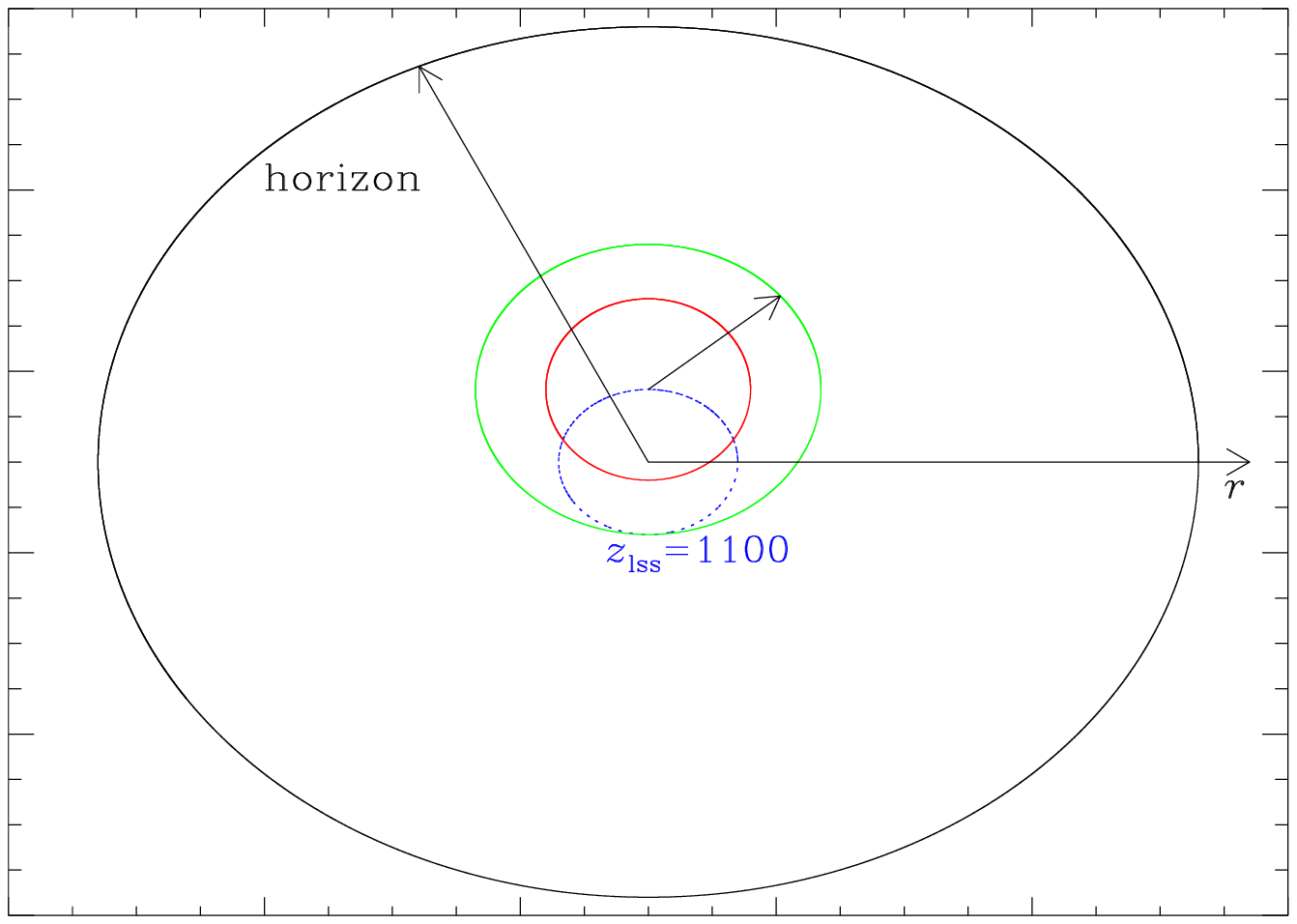}
\caption{Left panel: Sketch of the evolution of the horizon. The
origin of the coordinates is chosen to be Earth. The red circles
represent the size of the horizon at the time of equality, $z_{\rm eq}
\simeq 10^4$. The green circles represent the horizon at the time of
recombination $z_{\rm rec}\simeq 1100$. The black circle represents
the horizon today. The dotted blue circle represents the surface of
last scattering viewed from Earth. The angle $\Delta \Omega $ is the
angular size of the horizon at recombination viewed from Earth.  Right
panel: Sketch of the evolution of the horizon in an inflationary
universe. The conventions are the same as in the left panel. The
horizon at recombination now includes the last scattering surface and
there is no horizon problem anymore}
\label{hordiag}
\end{center}
\end{figure*}
As a consequence, one expects the last scattering surface to be made
of $\simeq 1^{\circ}$ patches whose physical properties are completely
different (let us remind that the angular diameter of the moon seen
from the Earth is $\simeq 0.5^{\circ}$). This is obviously not the
case: up to tiny fluctuations of order $\delta T/T\simeq 10^{-5}$, the
CMB radiation is extremely homogeneous and isotropic. This paradox is
called the horizon problem. A solution to this problem is to assume
that the initial conditions were identical in all the causally
disconnected patches but this seems very difficult to justify. Another
solution is to switch on the inflationary phase.  To significantly
modify the solid angle in Eq.~(\ref{solid}), the unknown fluid
responsible for inflation must have an equation of state such that
$\omega _{_{\rm X}}<-1/3$. Indeed, if $1+3\omega _{_{\rm X}}>0$, then
the argument of the exponential in Eq.~(\ref{solid}) is negative and
the correction coming from the phase driven by the unknown fluid
becomes negligible. On the other hand, if $1+3\omega _{_{\rm X}}<0$,
then the correction can be very important, depending of course on the
value of the number of e-folds $N$. Writing that the last scattering
surface looks very isotropic, that is to say $\Delta \Omega >4\pi $,
allows us to put a constraint on this quantity. One obtains
$N\agt -4+\ln z_{\rm end}$.
Notice that it is necessary to assume that $\omega _{_{\rm X}}$ is not
too close to $-1/3$ otherwise terms like $1+3\omega _{_{\rm X}}$, that
we have neglected, could also have an effect on the constraint derived
above.

\par

Let us now try to better understand and to physically interpret what
has been done. This is summarized in Fig.~\ref{hordiag} that we now
describe in more details. The proper distance to the last scattering
surface is given by
\begin{equation}
d_{\rm lss}=a_0\int _{t_{\rm lss}}^{t_0} \frac{{\rm d}\tau}{a(\tau)} 
=\frac{2}{H_0}\left[1-
\biggl(\frac{a_{\rm lss}}{a_0}\biggr)^{1/2}\right] \, .
\end{equation}
This is approximatively the Hubble distance today, defined by $\ell
_{_{\rm H}}\equiv H_0^{-1}$, since we have $d_{\rm lss}\simeq
2H_0^{-1}\simeq 6000 h^{-1}\mbox{Mpc} \simeq {\cal O}(1)\ell _{_{\rm
H}}$ where we have used $H_0\equiv 100h \mbox{km} \, \mbox{s}^{-1}
\mbox{Mpc}^{-1}$. Obviously, this number does not depend on the fact
that there is a phase of inflation or not. On the other hand, the size
of the horizon today is given by the following expression
\begin{widetext}
\begin{eqnarray}
\label{horizontoday}
d_{_{\rm H}}(t_0) =a_0
\int _{0}^{t_0} \frac{{\rm d}\tau}{a(\tau)} &=&
a_0\times 
\frac{1}{a_0H_0}\biggl(\frac{a_{\rm lss}}{a_0}\biggr)^{1/2} 
\left\{1+\frac{1-3\omega _{_{\rm X}}}{1+3\omega _{_{\rm X}}}
\frac{a_{\rm end}}{a_{\rm lss}}\left[1-
\left(\frac{a_{\rm i}}{a_{\rm end}}
\right)^{(1+3\omega _{_{\rm X}})/2}
\right]\right\} \nonumber \\
& & +a_0\times \frac{2}{a_0H_0}\left[1-
\left(\frac{a_{\rm lss}}{a_0}\right)^{1/2}\right]\, .
\end{eqnarray}
\end{widetext}
If there is no phase of inflation (or if $1+3\omega _{_{\rm X}}>0$)
then one has $d_{_{\rm H}}(t_0)\simeq 2H_0^{-1}\simeq d_{\rm
lss}\simeq \ell _{_{\rm H}}$. This is why, in the left panel in
Fig.~\ref{hordiag}, the (black) horizon and the (blue) last scattering
surface have about the same size. The horizon at recombination has
been calculated in Eq.~(\ref{horizonlss}). If there is no phase of
inflation then one has $d_{_{\rm H}}(t_{\rm lss})\simeq
H_0^{-1}(1+z_{\rm lss})^{-3/2} \ll d_{\rm lss}$.
This is why in the left panel in Fig.~\ref{hordiag}, the red and the
green circles are small in comparison with the blue circle
representing the last scattering surface. This is also the heart of
the horizon problem: without a phase of inflation, the horizon at
recombination is too small in comparison with the last scattering
surface and, as a consequence, its angular size is only $\simeq
1^{\circ}$ as we have calculated previously.

\par

Let us now turn to the inflationary solution and the right panel in
Fig.~\ref{hordiag}. The proper distance to the last scattering surface
is not modified. But, and this is the crucial point, the size of the
horizon is now completely different. Using Eq.~(\ref{horizonlss}), with
now $1+3\omega _{_{\rm X}}<0$, we obtain
\begin{equation}
d_{_{\rm H}}(t_{\rm lss})\simeq \frac{1}{H_0}(1+z_{\rm lss})^{-3/2}
\left[1+\frac{z_{\rm lss}}{z_{\rm end}}\left(\frac{a_{\rm end}}{a_{\rm i}}
\right)\right]\gg d_{\rm lss}\, .
\end{equation}
This is why in the right panel in Fig.~\ref{hordiag} the (green)
horizon now encompasses the (blue) last scattering surface. Another
consequence is that the (black) horizon today is now much bigger than
the Hubble scale which, as already mentioned, is still approximatively
equal to the size of the blue last scattering surface. For the purpose
of illustration let us take the example of chaotic inflation.  In this
case we have $z_{\rm end}\simeq 10^{28}$ and $a_{\rm end }/a_{\rm i}
\simeq \exp (10^{28})$ from which we deduce that (!)  $d_{_{\rm
H}}(t_0)\simeq 3\times 10^{43429421} h^{-1}\mbox{Mpc}$.
Clearly this scale is totally different from the Hubble scale and, in
the context of an inflationary universe, one should carefully make the
difference between those two scales.

\subsection{The flatness problem}

This problem becomes more apparent if the Friedman equation is cast
into a different form. Let us define the parameter $\Omega _{\rm i}$,
which gives the relative contribution of the fluid ``i'' to the total
amount of energy density present in the Universe, by $\Omega _{\rm
i}(t)\equiv \rho _{\rm i}(t)/\rho _{\rm cri}(t)$,
where the critical energy density is $\rho _{\rm cri}\equiv
3H^2/\kappa $. This last quantity is nothing but the total energy
density of a universe with flat space-like sections. This is a
time-dependent quantity. The Friedman equation takes the form
$k/(a^2H^2)=\sum _{i=1}^{N}\Omega _i(t)-1 \equiv \Omega _{_{\rm T}}(t)
-1$.
The parameter $\Omega _{_{\rm T}}(t)$ directly gives the sign of the
curvature of the space-like sections. Since $k$ is not a function of
time, the sign of $\Omega _{_{\rm T}}-1$ cannot change during the
cosmic evolution. In general, it is difficult to solve the
differential equation giving the time evolution of $\Omega _{_{\rm
T}}$(t) and to obtain the explicit time dependence of $\Omega _{_{\rm
T}}$. However, it is possible to express $\Omega _{_{\rm T}}$ in terms
of the scale factor $a(t)$, at least in the case where all the fluids
have a constant equation of state parameter. One obtains
\begin{widetext}
\begin{equation}
\Omega _{_{\rm T}}(a)=\sum _{i=1}^{N}\Omega _i(t_0)
\left(\frac{a}{a_0}\right)
^{-3(1+\omega _i)}\left\{\sum _{\rm j=1}^{N}\Omega _j(t_0)
\left(\frac{a}{a_0}\right)^{-3(1+\omega _j)}
-\left[\Omega _{_{\rm T}}(t_0)-1\right]\left(\frac{a}{a_0}\right)
^{-2}\right\}^{-1}\, .
\end{equation}
\end{widetext}
If one assumes that only radiation and matter are present then, as
$a/a_0$ goes to zero, it is clear that radiation becomes dominant. In
this case, a good approximation of the previous equation is
\begin{equation}
\label{ombb}
\Omega _{_{\rm T}}(t)-1  \simeq \frac{\Omega _{_{\rm T}}(t_0)-1}
{\Omega _{\rm rad}(t_0)}\biggl(\frac{a}{a_0}\biggr)^2 
=\frac{\Omega _{_{\rm T}}(t_0)-1}
{\Omega _{\rm rad}(t_0)}\biggl(\frac{1}{z+1}\biggr)^2 \, .
\end{equation}
Today it is known that $\vert \Omega _{_{\rm T}}(t_0)-1\vert
<0.1$. This clearly means that, at high redshifts, the quantity $\vert
\Omega _{_{\rm T}}(z)-1\vert $ was extremely close to zero. For
instance, at the redshift of nucleosynthesis, $z_{\rm nuc}\simeq
3\times 10^8$, one has $\vert \Omega _{_{\rm T}}(z_{\rm nuc})-1\vert
\simeq {\cal O}(10^{-14})$ where we have taken $\Omega _{\rm
rad}(t_0)\simeq 10^{-4}$. It is difficult to understand why this
quantity was so fine-tuned in the early Universe. At Grand Unified
Theory (GUT) scale ($z_{_{\rm GUT}}\simeq 10^{28}$), the constraint
becomes even worse $\vert \Omega _{_{\rm T}}(z_{_{\rm GUT}})-1\vert
\simeq {\cal O}(10^{-52})$. To explain this fact, we have two possible
solutions: (i) we simply assume that the initial conditions were
fine-tuned in the early Universe or (ii) we find a mechanism which
automatically produces such a small value at high redshifts. Since, as
already mentioned for the horizon problem, the first explanation seems
artificial, let us concentrate on the second one. Thus, we assume
that, for redshifts $z>z_{\rm end}$, the Universe was dominated by
another type of matter, different from matter or radiation. We simply
characterize this unknown fluid $X$ by its equation of state $\omega
_{_{\rm X}}$.  The equation of state should be chosen such that, from
any reasonable (i.e. not fine-tuned) initial conditions in the very
early Universe $z\gg z_{\rm end}$, it automatically produces a $\Omega
_{_{\rm T}}-1$ close to zero, with the required accuracy at $z=z_{\rm
end}$. $\Omega _{_{\rm T}}$ can be written as
\begin{equation}
\label{ominf}
\Omega _{_{\rm T}}(a)=\frac{\Omega _{_{\rm X}}(a_{\rm i})}
{\Omega _{_{\rm X}}(a_{\rm i})+
\left[1-\Omega _{_{\rm T}}(a_{\rm i})\right]
\left(\displaystyle{\frac{a}{a_{\rm i}}}\right)^
{1+3\omega _{_{\rm X}}}} \,,
\end{equation}
where $a_{\rm i}$ is the value of the scale factor at some initial
redshift $z_{\rm i}$. Since $a_{\rm end}/a_{\rm i}\gg 1$, the
condition $\Omega _{_{\rm T}}(a_{\rm end})\simeq 1$ is clearly
equivalent to $1+3\omega _{_{\rm X}}<0$. Then, from any initial
conditions at $z=z_{\rm i}$, the value of $\Omega _{_{\rm T}}(a_{\rm
end})$ will be pushed toward one as long as $X$ dominates. Therefore,
one recovers the fact that a fluid with a negative equation of state
parameter can solve a problem of the hot Big Bang model. One can even
derive the constraint that the parameters describing the epoch
dominated by the fluid $X$ must satisfy. If one requires that $\Omega
_{_{\rm T}}$ has been pushed so close to one during the phase
dominated by $X$ that the remaining difference $\Omega _{_{\rm T}}-1$
will not sufficiently increased during the radiation and dominated
epochs to compensate the first effect and to be distinguishable from
zero today, one arrives at [from Eqs.~(\ref{ombb}) and (\ref{ominf})
written at $z=z_{\rm end}$]
\begin{equation}
\biggl(\frac{a_{\rm end}}{a_{\rm i}}\biggr)^{1+3\omega _{_{\rm X}}}
={\rm e}^{N\left(1+3\omega _{_{\rm X}}\right)} 
\lta 10^4 \times z_{\rm end}^{-2}\, ,
\end{equation}
which can also be expressed as $N \gta -4 +\ln z_{\rm end}$,
where we have assumed for simplicity that $\vert 1+3\omega _{_{\rm
X}}\vert ={\cal O}(1)$. It is quite remarkable that this constraint be
the same as the one derived from the requirement that the horizon
problem is solved. To conclude, let us give some numerical examples:
for $z_{\rm end}\simeq 10^{10}$, {\it i.e.} two orders of magnitude
above nucleosynthesis, one has $a_{\rm end}/a_{\rm i}\simeq 10^8$ that
is to say $\simeq 19$ e-foldings. For $z_{\rm end}\simeq z_{_{\rm
GUT}}$, one obtains $a_{\rm end}/a_{\rm i}\simeq 10^{26}$, namely
$\simeq 60$ e-foldings.

\par

The main lesson of the previous calculations is that, assuming an
epoch in the early Universe dominated by a fluid the equation of state
parameter of which is negative, provides an elegant way to solve the
problems of the standard hot Big Bang model. Here, the important point
is that the detailed properties of the unknown fluid and/or its
physical nature are unimportant, at least at the background level,
provided the equation of state parameter is negative. This makes the
inflationary solution quite generic.

\subsection{Single scalar field inflation}

We have seen in the previous sections that inflation can be caused by
any fluid such that $\rho +3p<0$. We now discuss a concrete
realization of the inflationary mechanism. Inflation is supposed to
take place in the very early universe, at very high energies. At those
scales, the fluid description of matter is not expected to hold
anymore and (quantum) field theory seems to be the most appropriate
way to describe the behavior of matter. The simplest example,
compatible with the symmetries of the FLRW metric, is a scalar field
$\phi _\zero(\eta )$. This field will be called the inflaton in what
follows. The corresponding Lagrangian reads
\begin{equation}
S=-\int {\rm d}^4x\sqrt{-g}
\biggl[\frac{1}{2}g^{\mu \nu}\partial
_{\mu} \phi _\zero\partial _{\nu}\phi _\zero+V(\phi _\zero)\biggr]\, ,
\end{equation}
where $V(\phi _\zero )$ is the potential, {\it a priori} a free
function but we will see that, in order to have a successful
inflationary phase, its shape must satisfy some constraints. The
stress-energy tensor can be written as
\begin{equation}
T_{\mu \nu}=\partial _{\mu }\phi _\zero\partial _{\nu }\phi _\zero
-g_{\mu \nu}\biggl[\frac{1}{2}g^{\alpha \beta }
\partial _{\alpha }\phi _\zero\partial _{\beta }\phi _\zero
+V(\phi _\zero)\biggr]\, .
\end{equation}
 From this expression, one sees that the scalar field can also be
viewed as a perfect fluid. The energy density and the pressure are
defined according to $T^0{}_0=-\rho $ and $T^i{}_j=p\delta ^i{}_j$ and
read
\begin{equation}
\label{rhopsf}
\rho = \frac{1}{2}\frac{(\phi _\zero')^2}{a^2}+V(\phi _\zero), \quad 
p= \frac{1}{2}\frac{(\phi _\zero')^2}{a^2}-V(\phi _\zero)\, .
\end{equation}
The conservation equation can be obtained by inserting the previous
expressions of the energy density and pressure into
Eq.~(\ref{conservation}). Assuming $\phi _\zero'\neq 0$, this
reproduces the Klein-Gordon equation written in a FLRW background,
namely
\begin{equation}
\phi _\zero''+2\frac{a'}{a}\phi _\zero'
+a^2\frac{{\rm d}V(\phi _\zero)}{{\rm d}\phi _\zero}=0\, .
\end{equation}
As already mentioned before, the other equation of conservation
expresses the fact that the scalar field is homogeneous and therefore
does not bring any new information. Finally, a comment is in order
about the equation of state.  In general, there is no simple link
between $\rho $ and $p$ except when the kinetic energy dominates the
potential energy where $\omega \simeq 1$, {\it i.e.} the case of stiff
matter or, on the contrary, when the potential energy dominates the
kinetic energy for which one obtains $\omega \simeq -1$. This last
case is of course the most interesting for our purpose. This shows
that inflation corresponds to a regime where the potential energy
dominates the kinetic energy: $V(\phi _\zero )\gg (\phi _\zero ')^2$, 
see Eqs.~(\ref{rhopsf}). We also note in passing that an equation of
state $p\simeq -\rho $ implies that the energy density of the field
will be almost constant during inflation. The fact that the kinetic
energy is small during inflation means that the potential should be
very flat which is the main requirement for a successful model of
inflation if this one is caused by a scalar field.

\par

In general, the equations of motion can be integrated exactly only for
a very restricted class of potentials. On the contrary, one would like
to able to characterize this motion for any given, sufficiently flat,
potentials. To reach this goal, we clearly need a scheme of
approximation. Since the kinetic energy to potential energy ratio and
the scalar field acceleration to the scalar field velocity ratio are
small, this suggests to view these two ratios as parameters in which a
systematic expansion is performed. The slow-roll motion of the scalar
field is controlled by the three ``slow-roll parameters'' (at leading
order, see e.g.~Ref.~\cite{SL}) defined by:
\begin{widetext}
\begin{equation}
\label{defepsilon}
\epsilon \equiv 3 \frac{\dot{\phi_\zero}^2}{2} 
\left(\frac{\dot{\phi _\zero}^2}{2} + V\right)^{-1} 
= -\frac{\dot{H}}{H^2}=1-\frac{{\cal H}'}{{\cal H}^2}\, ,
\quad
\delta \equiv -\frac{\ddot{\phi_\zero}}{H\dot{\phi_\zero }} =
- \frac{\dot{\epsilon }}{2 H \epsilon }+\epsilon\ , \quad 
\xi \equiv \frac{\dot{\epsilon }-\dot{\delta }}{H}\ .
\end{equation}
\end{widetext}
Some remarks are in order at this point. First of all, we have
introduced a third slow-roll parameters, $\xi $. This is necessary if
one wants to establish the exact equations of motion of $\epsilon $
and $\delta $. Secondly, the slow-roll conditions are satisfied if
$\epsilon$ and $\delta$ are much smaller than one and if $\xi = {\cal
O}(\epsilon^2,\delta^2,\epsilon\delta)$. Since the equations of motion
for $\epsilon$ and $\delta$ can be written as:
\begin{equation}
\label{eqmotionsrpara}
\frac{\dot{\epsilon}}{H}=2\epsilon (\epsilon -\delta)\ , \quad 
\frac{\dot{\delta }}{H}=2\epsilon (\epsilon -\delta )-\xi \, ,
\end{equation}
it is clear that this amounts to consider $\epsilon$ and $\delta$ as
constants. This property turns out to be crucial for the calculation
of the perturbations. Thirdly, inflation stops when $\epsilon
=-\dot{H}/H^2=1$. Finally, it is also convenient to re-express the
slow-roll parameters in terms of the inflaton potential. One can show
that
\begin{equation}
\label{srpotential}
\epsilon \simeq \frac{m_{_{\rm Pl}}^2}{16\pi}
\biggl(\frac{V'}{V}\biggr)^2, \quad 
\delta \simeq -\frac{m_{_{\rm Pl}}^2}{16\pi}
\biggl(\frac{V'}{V}\biggr)^2+
\frac{m_{_{\rm Pl}}^2}{8\pi }\frac{V''}{V}\, ,
\end{equation}
where, here, a prime means a derivative with respect to the scalar
field (as expected, the third slow roll parameter involves the third
derivative of the potential). This suggests a new interpretation of
the slow-roll approximation: the slow-roll parameters controls the
deviation of the inflaton potential from perfect flatness (the case of
a cosmological constant) and hence are given by the successive field
derivatives of the potential.

\par

Yet another way to see the slow-roll approximation is the
following. The perfect slow-low regime is when the inflaton potential
is a constant, {\it i.e.}  is exactly flat. In this case, one has
$\epsilon =\delta =0$ and the corresponding solution of the Einstein
equations is the de Sitter space-time with the scale factor $a(\eta
)\propto \vert \eta \vert ^{-1}$. Somehow, the slow-roll approximation
is an expansion around this solution. To illustrate this point, let us
consider the exact equation:
\begin{equation}
\label{etasr}
\eta = - \int 
\frac{1}{1-\epsilon}{\rm d}\biggl(\frac{1}{{\cal H}}\biggr)\, ,
\end{equation}
which comes directly from the definition of $\epsilon \equiv 1-{\cal
H}'/{\cal H}^2$ written in terms of the conformal time. An integration
by parts and the use of the equation of motion of the slow-roll
parameter $\epsilon $ allows us to reduce the previous equation to
\begin{equation}
\eta =-\frac{1}{(1-\epsilon ){\cal H}}-\int \frac{2\epsilon (\epsilon
-\delta )}{(1-\epsilon )^3}{\rm d}\biggl(\frac{1}{{\cal
H}}\biggr)\, .
\end{equation}
So far, no approximation has been made. At leading order, $\epsilon$
is a constant and the previous equation reduces to $aH \approx
-(1+\epsilon)/\eta$. This is equivalent to a scale factor which
behaves as $a(\eta)\approx \ell_\zero |\eta|^{-1-\epsilon}$. Therefore,
the slow-roll approximation consists in slightly modifying the de
Sitter expansion by changing the power index in the expression of the
scale factor. Interestingly enough, the effective power index (at
leading order) only depends on $\epsilon$. We will see that the second
slow-roll parameters will show up in the calculation when we consider 
inflationary cosmological perturbations.

\par

Finally, it is also useful to make use of the set of horizon flow
functions, first introduced in Ref.~\cite{flow}. The big advantage of
these parameters is that there are defined in terms of the scale
factor only and thus do not rely on the fact that inflation is caused
by one scalar field. In particular, these parameters could still be
used in a multi-fields model of inflation whereas the set introduced
previously should be modified. The zeroth horizon flow function is
defined by $\epsilon_\zero \equiv H(N_{\rm i})/H(N)$, where $N$ is the
number of e-folds after an arbitrary initial time. The hierarchy of
horizon flow functions is then defined according to
\begin{equation}
\label{flow}
\epsilon _{n+1}\equiv 
\frac{{\rm d}\ln \vert \epsilon _n\vert }{{\rm d}N}, \quad 
n\ge 0\, .
\end{equation}
The link between the horizon flow functions and the set $\{\epsilon ,
\delta , \xi \}$ can be expressed as
\begin{equation}
\epsilon =\epsilon _1\, ,\quad 
\delta =\epsilon _1-\frac{1}{2}\epsilon _2\, ,
\quad
\xi =\frac{1}{2}\epsilon _2 \epsilon _3 \, .
\end{equation}
The fact that $\xi $ is of higher order than the two first slow-roll
parameters is now obvious which is another advantage of the horizon
flow parameters.

\section{Inflationary cosmological perturbations}

\subsection{Gauge-invariant formalism}

The perturbed line element can be written as \cite{MFB}:
\begin{eqnarray}
\label{metricgi}
{\rm d}s^2 &=& 
a^2(\eta )\{-(1-2\phi ){\rm d}\eta ^2+2({\rm \partial}_iB){\rm d}x^i
{\rm d}\eta \nonumber \\ & & +[(1-2\psi )\delta _{ij}
+2{\rm \partial }_i{\rm \partial }_jE+h_{ij}]{\rm d}x^i{\rm d}x^j\}\ .
\end{eqnarray}
In the above metric, the functions $\phi $, $B$, $\psi $ and $E$
represent the scalar sector whereas the tensor $h_{ij}$, satisfying
$h_i{}^i=h_{ij}{}^{,j}=0$, represents the gravitational waves. There
are no vector perturbations because a single scalar field cannot seed
rotational perturbations. At the linear level, the two types of
perturbations decouple and thus can be treated separately.

\par

The scalar sector suffers from the gauge problem. This means that an
infinitesimal transformation of coordinates ({\it i.e.} a ``gauge
transformation'') could mimic a physical deformation of the underlying
background space-time and thus could be confused with a physical mode
of perturbations. In order to deal with this problem and to retain
only the physical modes, one can either fix the gauge or work with
gauge invariant quantities. Here, we choose the latter
solution. Scalar perturbations of the geometry can be characterized by
the gauge-invariant Bardeen potentials $\Phi $~\cite{B1} and
fluctuations in the scalar field are characterized by the
gauge-invariant quantity $\delta\phi ^{\rm (gi)}$
\begin{eqnarray}
\Phi &=& \phi +\frac{1}{a}\left[a\left(B-E'\right)\right]'\, ,
\\ 
\delta \phi ^{\rm (gi)} &=& \delta \phi 
+\phi _{\zero }'\left(B-E'\right)\, .
\end{eqnarray}
We have two gauge invariant quantities but only one degree of freedom
since $\Phi $ and $\delta \phi ^{\rm (gi)}$ are linked by the
perturbed Einstein equations. As a consequence, if $\phi _\zero'\neq
0$, then the whole problem can be reduced to the study of a single
gauge-invariant variable (the so-called Mukhanov-Sasaki variable)
defined by \cite{MuChi}
\begin{equation}
v\equiv a\left[\delta \phi^{\rm (gi)}+ \phi _\zero'\frac{\Phi }{\calH}
\right]\, .
\end{equation}
In fact, it turns out to be more convenient to work with the rescaled
variable $\mu _{_{\rm S}}$ defined by $\mu _{_{\rm S}}\equiv
-\sqrt{2\kappa }v$. Density perturbations are also often characterized
by the so-called conserved quantity $\zeta $~\cite{L,MS} defined by
\begin{equation}
\zeta \equiv \frac23\frac{{\cal H}^{-1}\Phi '+\Phi
}{1+\omega }+\Phi \, . 
\end{equation}
The quantity $\muS$ is related to $\zeta$ by $\muS=-2a\sqrt{\gamma
}\zeta $, where $\gamma =1-{\calH}'/{\calH}^2$. The background
function $\gamma $ reduces to a constant, $(2+\beta )/(1+\beta )$, for
power-law scale factors $a(\eta ) \propto (-\eta )^{1+\beta }$. In
particular, it is zero for the de Sitter space-time since, in this
case, $\beta =-2$. The equation of motion of the quantity $\muS$
reads~\cite{MFB}
\begin{equation}
\label{eomscalar}
\muS''+\left[k^2-
\frac{(a\sqrt{\gamma})''}{(a\sqrt{\gamma })}\right]\muS=0\, ,
\end{equation}
where $k$ is the co-moving wavenumber of the corresponding Fourier
mode. This equation is similar to a Schr\"odinger time-independent
equation where the usual role of the radial coordinate is now played
by the conformal time (this is why the name ``time-independent
Schr\"odinger equation'' is particularly unfortunate in the present
context!). The effective potential $U_{_{\rm S}}\equiv
(a\sqrt{\gamma})''/(a\sqrt{\gamma })$ involves the scale factor
$a(\eta )$ and its derivative (up to the fourth order) only.
 
\par

In the tensor sector (which is gauge invariant by definition) we
define the quantity $\muT$ for each mode $k$ according to
$h_{ij}=(\muT/a)Q_{ij}$, where $Q_{ij}$ are the (transverse and
traceless) eigentensors of the Laplace operator on the space-like
sections. The equation of motion of $\mu _{_{\rm T}}$ is given by
\cite{Gpara}:
\begin{equation}
\label{eomtensor}
\muT''+\left(k^2-\frac{a''}{a}\right)\muT=0\,  .
\end{equation} 
This formula is similar to the equation of motion of density
perturbations. The only difference is that the effective potential,
$U_{_{\rm T}}=a''/a$, now involves the derivatives of the scale factor
only up to the second order.

\par

Therefore, we have shown that both types of perturbations obey the
same type of equation of motion. The ``time-independent
Schr\"odinger'' equation can also be viewed as the equation of motion
of an harmonic oscillator whose frequency explicitly depends on time,
namely the equation of a parametric oscillator~\cite{MS}
\begin{equation}
\label{eq:evol}
\muST''+\omegaST^2(k,\eta ) \muST=0 ,
\end{equation}
with $\omegaS^2=k^2 -(a\sqrt{\gamma })''/(a\sqrt{\gamma })$,
$\omegaT^2=k^2 -a''/a$.

\par

Finally, the mode functions $\mu _{_{\rm S,T}}$ are quantities of
interest because the power spectra of density perturbations and
gravitational waves, which are observables, directly involve
them. Explicitly, one has
\begin{equation}
\label{spec}
k^3P_{\zeta }(k)=\frac{k^3}{8\pi ^2}\biggl\vert \frac{\mu
_{_{\rm S}}}{a\sqrt{\gamma }}\biggr\vert ^2 , \quad 
k^3P_h(k)=\frac{2k^3}{\pi ^2}\biggl \vert \frac{\mu _{_{\rm T}}}{a}
\biggr \vert ^2 .
\end{equation}
The spectral indices and their running are defined by the coefficients
of Taylor expansions of the power spectra with respect to $\ln k$,
evaluated at an arbitrary pivot scale $k_*$.
\begin{equation}
\label{n}
n_{_{\rm S}} -1 \equiv {{\rm d} \ln {\cal P}_\zeta\over {\rm d} \ln k}
\biggr\vert_{k=k_*},
\quad
n_{_{\rm T}} \equiv {{\rm d} \ln {\cal P}_h\over {\rm d} \ln k}
\biggr\vert_{k=k_*} \, ,
\end{equation}
are the spectral indices. The fact that scale invariance corresponds
to $n_{_{\rm S}}=1$ for density perturbations and to $n_{_{\rm T}}=0$
for gravitational waves has no deep meaning and is just an historical
accident. The two following expressions
\begin{equation}
\label{alpha}
\alpha_{_{\rm S}} \equiv {{\rm d}^2 \ln {\cal P}_\zeta\over 
{\rm d} (\ln k)^2}\biggr\vert_{k=k_*}, 
\quad
\alpha_{_{\rm T}} \equiv  {{\rm d}^2 \ln {\cal P}_h\over {\rm d}( \ln k)^2}
\biggr\vert_{k=k_*}\, , 
\end{equation}
define the ``running'' of these indices. In principle, we could also
define the running of the running and so on.

\par

In order to compute $k^3P_{\zeta }(k)$ and $k^3P_h(k)$, one must
integrate the equation of motion (\ref{eq:evol}) and specify what the
initial conditions are.  We now turn to these questions.

\subsection{Qualitative behavior of the solutions}

The advantage of the previous formulation is that it allows to guess
the form of the solutions very easily. It is essentially determined by
three scales. Firstly, one has the physical wavelength of a given
Fourier mode $\lambda (\eta )=(2\pi /k)a(\eta )$.
A second length scale important for the problem is given by the
effective potential. To be specific, one has $\ell _{_{\rm U}}(\eta )
=a(\eta )/\sqrt{U_{_{\rm S,T}}(\eta )}$.
Finally, a third scale is the Hubble scale whose definition reads
$\ell _{_{\rm H}}\equiv a^2/a'$. A priori, the Hubble scale and 
the potential scale $\ell _{_{\rm U}}$ are different.

\par

Let us now investigate how these scales behave in a typical model of
slow-roll inflation. For the purpose of illustration, we only consider
the scale factor $a(\eta )\propto \vert \eta \vert ^{-1}$ during
inflation since we have seen before that any model of slow-roll
inflation can be seen as a small deformation of the de Sitter
space-time.
\begin{figure}[t]
\hspace*{-0.9cm}
\includegraphics*[width=10cm, height=8cm, angle=0]{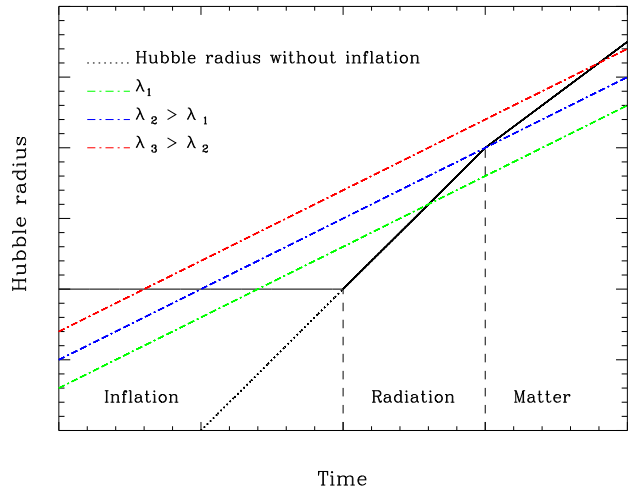}
\caption{Evolution of the Hubble radius and of three physical
wavelengths with different comoving wavenumbers during the
inflationary phase and the subsequent radiation and matter dominated
epochs. Without inflation, the wavelengths of the mode are
super-Hubble initially whereas in the case where inflation takes
place, they are sub-Hubble which permits to set up sensible initial
conditions.}
\label{scaleinf}
\end{figure}
One immediately obtains that the Hubble radius is constant during
inflation while it varies as $\propto a^2$ during the radiation era
and as $\propto a^{3/2}$ during the matter dominated era. Initially,
the physical wavelengths are therefore smaller than the Hubble radius
(in principle even smaller than the Planck length, see below) but
because of the inflationary expansion of the background they become,
at some point, larger than the Hubble radius. The time at which a
Fourier mode exits the Hubble radius depends on the co-moving
wavenumber of the corresponding mode. They will re-enter the Hubble
radius later on, either during the radiation or matter dominated
epochs because the Hubble radius behave differently during those
eras. It is worth noticing that, without a phase of inflation, the
modes would have always been outside the Hubble radius. The fact that
there is a regime where the modes are sub-Hubble is therefore a
specific feature of the inflationary background and plays a crucial
role in our ability to fix well-defined initial conditions. The
evolution of a Fourier mode is represented in Fig.~\ref{scaleinf}.

\par

It is also interesting to remark than the physical wavelengths are
always inside the horizon which they never exit. It is therefore
mandatory to distinguish the horizon from the Hubble scale. In fact it
is possible to prove that, as soon as a scale is inside the horizon,
it will remain so for ever. This is simply due to the fact that the
ratio of the horizon to the physical scale at time $t$ is given by
\begin{equation}
\frac{d_{_{\rm H}}}{\lambda }=\frac{k}{2\pi }
\int _{t_{\rm i}}^{t_0}\frac{{\rm d}\tau }{a(\tau )}
+
\frac{k}{2\pi }
\int _{t_0}^{t}\frac{{\rm d}\tau }{a(\tau )}\, ,
\end{equation}
where $k$ is the co-moving wavenumber of the scale under
consideration.  The first term is by assumption greater than one and
the second one is positive, hence the above-mentioned statement

\par


Looking at the equation of motion, one sees that, {\it a priori}, the
behavior of the solution is not controlled by the Hubble scale as
often said in the literature (sometimes, it is also claimed that the
horizon determines the qualitative behavior of the solutions!) but by
the scale $\ell _{_{\rm U}}$, {\it i.e.} by the shape of the effective
potential. However, it turns out that, for slow-roll inflation (in
fact for power-law inflation), the behaviors of $\ell _{_{\rm U}}$
and $\ell _{_{\rm H}}$ are similar and, therefore, the concepts of
Hubble and potential scales can be used almost interchangeably in this
situation. This is not the case in general. For instance, this is
incorrect in a bouncing universe, see Ref.~\cite{bounce}.

\par

Let us now study in more details the shape of the effective
potential. The function $\gamma $ is a constant for power law scale
factors and, as a consequence, the two types of perturbations acquire
the same potential during inflation, namely $U_{_{\rm S,T}}(\eta
)\simeq \eta ^{-2} \simeq {\cal H}^2$. During the radiation dominated
epoch, the scale factor behaves as $a(\eta )\propto \eta $ and
therefore the effective potential vanishes. As a consequence, the
typical form of the potential is given by the solid line in
Fig.~\ref{potdp}. The exact form of the potential during the
transition ({\it i.e.} during the reheating) is very complicated and
has not been taken into account here. A smooth interpolation has been
assumed between the two eras. The two regimes described before
re-appear in a different way. Initially, a given mode is above the
barrier. This corresponds to the case where the mode is
sub-Hubble. Then, due to the inflationary evolution, it crosses the
potential at some (scale dependent) time and becomes
``sub-potential''. This corresponds to a super-Hubble mode. The times
of Hubble and potential crossings are not the same but, as already
mentioned, in the case of slow-roll inflation they are of the same
order of magnitude. Therefore, in this case, we have the approximate
correspondence ``inside the Hubble scale/above the potential'' and
``outside the Hubble scale/below the potential''. Let us notice that
this is valid as long as the details of the reheating process only
modify the shape of the potential such that the modes of interest
always remain below the potential during the transition from inflation
to radiation.

\par

The two previous regimes correspond to two types of solution. In 
the first regime, $k^2\gg U(\eta )$, and the mode function 
oscillates, 
\begin{equation}
\label{subhubble}
\mu _{_{\rm S,T}}\simeq A_1(k){\rm e}^{-ik\eta }
+A_2(k){\rm e}^{ik\eta }\, .
\end{equation}
On the contrary, when the potential dominates, $k^2\ll U(\eta )$, the
solutions are of the form 
\begin{equation}
\label{superhubble}
\mu _{_{\rm S}}\simeq C_1(k)a\sqrt{\gamma }+C_2(k)a\sqrt{\gamma }
\int ^{\eta }\frac{{\rm d}\tau}{(a^2\gamma )(\tau )}\, ,
\end{equation}
and possess a growing and a decaying modes. For gravitational waves,
the solutions are the same except that one should take $\gamma =1$ in
the previous equation. It is interesting to notice that the previous
solutions are general and do not depend on the specific form of the
scale factor.

\par

The only thing which remains to be discussed are the initial
conditions, {\it i.e.} the choice of the coefficients $A_1(k)$ and
$A_2(k)$.

\subsection{WKB approximation and the initial conditions}

It has been established before that the mode functions $\muS $ and
$\muT$ obey the equation of a parametric oscillator. This strongly
suggests to use the WKB approximation to study the solutions of this
equation~\cite{MSwkb}.  For this purpose, let us define the WKB mode function,
$\mu_{_{\rm WKB}}$, by the following expression
\begin{equation}
\label{wkbsol}
\displaystyle \mu_{_{\rm WKB}}(k,\eta )\equiv 
\frac{1}{\sqrt{2\omega(k,\eta)}}
{\rm e}^{\displaystyle \pm i\int^{\eta }\omega(k,\tau){\rm d}\tau}\, .
\end{equation}
The mode function $\mu_{_{\rm WKB}}$ represents the leading order term
of a semi-classical expansion, {\it i.e.} it is only an approximation
to the actual solution of Eq.~(\ref{eq:evol}). This can also be viewed
from the fact that $\mu _{_{\rm WKB}}$ satisfies the following
differential equation
\begin{equation}
\label{eqwkb}
\mu _{_{\rm WKB}}''(k,\eta ) +\left[\omega ^2(k,\eta )-
Q(k,\eta )\right]\mu_{_{\rm WKB}}(k,\eta )=0\, ,
\end{equation}
which is not similar to Eq.~(\ref{eq:evol}). In the above formula, the
quantity $Q(k,\eta )$ is given by
\begin{equation}
\label{defQ}
Q(k,\eta )\equiv 
\frac{3}{4}\frac{(\omega ')^2}{\omega ^2}
-\frac{\omega ''}{2\omega }\, ,
\end{equation}
and only depends on the time dependent frequency $\omega (k,\eta
)$. From Eqs.~(\ref{eq:evol}) and (\ref{eqwkb}), it is clear that the
mode function $\mu _{_{\rm WKB}}(k,\eta )$ is a good approximation of
the actual mode function $\mu (k,\eta )$ if the following condition is
satisfied: $\vert Q/\omega ^2\vert \ll 1$.
If, for simplicity, we only keep the first term in the expression
giving $Q(k,\eta )$, see Eq.~(\ref{defQ}), the above equation can also
be re-written under the more traditional form $({\rm d}U/{\rm d}\eta
)/(k^2-U)^{3/2}\ll 1$, which expresses the fact that the WKB
approximation breaks down at the turning point and is valid when the
potential does not vary too rapidly.

\par

Let us now test this criterion for the two regimes described
before. In the case of slow-roll inflation, we will see that the
effective potential, either for density perturbations or gravitational
waves, is of the form ${\cal O}(1)/\eta ^2$. Therefore, on sub-Hubble
scales, in the limit $\vert \eta \vert \rightarrow +\infty $, one has
$\omega \simeq k$, which implies $Q\simeq 0$ and therefore the $\vert
Q/\omega ^2\vert \ll 1$ is satisfied. On the contrary, on super-Hubble
scales, {\it i.e.}, in the limit $\vert \eta \vert \rightarrow 0$, one
has $\vert Q/\omega ^2\vert _{_{\rm S}} \simeq \vert Q/\omega ^2
\vert _{_{\rm T}}={\cal O}(1)$. 
Thus, the WKB approximation is not a good approximation in this
regime. 

\par

The fact that the WKB approximation works in the limit $\vert \eta
\vert \rightarrow +\infty $ allows us to fix well-motivated initial
conditions and is the reason why the inflationary mechanism for
structure formation is so attractive. Indeed, within the framework
described before, the natural choice is to take the adiabatic vacuum
as the initial state. Since, on sub-Hubble scales, $\omega
(k)\rightarrow k$, Eq.~(\ref{wkbsol}) implies that this corresponds to
coefficients $A_1(k)$ and $A_2(k)$ in Eq.~(\ref{subhubble}) such that
\begin{equation}
A_1(k)\propto \frac{1}{\sqrt{2k}}\, ,\quad A_2(k)=0\, .
\end{equation}
This completely fixes the initial conditions and allows us to
calculate the power spectrum unambiguously.

\par

Before turning to this calculation, let us quickly come back to the
fact that the WKB approximation breaks down on super-Hubble scales. In
fact, this problem bears a close resemblance with a situation
discussed by atomic physicists at the time quantum mechanics was
born. The subject debated was the application of the WKB approximation
to the motion in a central field of force and, more specifically, how
the Balmer formula, for the energy levels of hydrogenic atoms, can be
recovered within the WKB approximation. The effective frequency for
hydrogenic atoms is given by (obviously, in the atomic physics
context, the wave equation is not a differential equation with respect
to time but to the radial coordinate $r$)
\begin{equation}
\label{omeH}
\omega ^2(E,r)=\frac{2m}{\hbar^2}\biggl(E+\frac{Ze^2}{r}\biggr)
-\frac{\ell (\ell +1)}{r^2}\, ,
\end{equation}
where $Ze$ is the (attractive) central charge and $\ell$ the quantum
number of angular momentum. The symbol $E$ denotes the energy of the
particle and is negative in the case of a bound state. Apart from the
term $Ze^2/r$ and up to the identification $r \leftrightarrow \eta $,
the effective frequency has exactly the same form as $\omega _{_{\rm
S,T}}(k,\eta )$ during inflation. Therefore, calculating the evolution
of cosmological perturbations on super-Hubble scales ({\it i.e.}
$\vert \eta \vert \rightarrow 0$) is similar to determining the
behavior of the hydrogen atom wave function in the vicinity of the
nucleus ({\it i.e.} $r\rightarrow 0$). The calculation of the energy
levels by means of the WKB approximation was first addressed by
Kramers \cite{Kramers} and by Young and Uhlenbeck \cite{YU}. They
noticed that the Balmer formula was not properly recovered but did not
realize that this was due to a misuse of the WKB approximation. In
$1937$ the problem was considered again by Langer~\cite{Langer}. In a
remarkable article, he showed that the WKB approximation breaks down
at small $r$, for an effective frequency given by Eq.~(\ref{omeH})
and, in addition, he suggested a method to circumvent this
difficulty. Recently, this method has been applied to the calculation
of the cosmological perturbations in Ref.~\cite{MSwkb}. This gives rise
to a new method of approximation, different from the more traditional
slow-roll approximation. 

\subsection{Simple calculation of the inflationary power spectrum}

\begin{figure}[t]
\hspace*{-1.25cm}
\includegraphics*[width=10cm, height=8cm, angle=0]{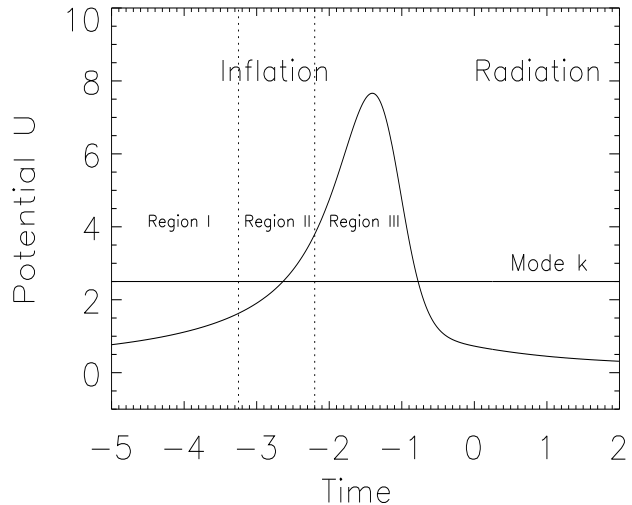}
\caption{Sketch of the effective potential of Eqs.~(\ref{eomscalar})
and (\ref{eomtensor}). During the inflationary phase the effective
potential behaves as $U\simeq \eta ^{-2}$ while during the radiation
dominated era it goes to zero. A smooth transition between these two
epochs has been assumed which does not take into account the details
of the reheating (and preheating) process.}
\label{potdp}
\end{figure}

Let us now evaluate $k^3P_{\zeta }$ in a very simple way in order to
understand why inflation leads to a scale invariant power spectrum.
On super-Hubble scale, {\it i.e.} in region III in Fig.~\ref{potdp},
the growing mode is given by $\muS \simeq C_1(k)a\sqrt{\gamma }$, see
Eq.~(\ref{superhubble}). Inserting this expression into
Eq.~(\ref{spec}), one obtains
\begin{equation}
\label{quickps}
k^3P_{\zeta }\propto k^3\vert C_1(k)\vert ^2 \, .
\end{equation} 
The next step is to relate the constant $C_1(k)$ to the initial
conditions, {\it i.e.} to $A_1(k)$. This can be done by writing the
continuity of the mode function $\muS$ at the time of potential
crossing, {\it i.e.} at the time where $k^2=U[\eta _{\rm j}(k)]$. 
In this case, one does not consider the details of region II in
Fig.~\ref{potdp}. One just matches by brute force the solutions of
regions I and III. This gives
\begin{equation}
A_1(k){\rm e}^{-ik\eta _{\rm j}}
=C_1(k)(a\sqrt{\gamma })(\eta _{\rm j})\, .
\end{equation}
In order to calculate the function $\eta _{\rm j}(k)$, one needs to
assume something about the scale factor.  Here, we consider power-law
inflation with the scale factor: $a(\eta )\propto \vert \eta \vert
^{1+\beta }$. In this case the effective potential is given by $U(\eta
)\propto \eta ^{-2}$ (and in fact the equation of motion can be
integrated exactly in terms of Bessel function) which amounts to take
$\eta _{\rm j}(k)\propto k^{-1}$.  Using also the fact that $\gamma $
is a constant for power-law inflation, the constant $C_1(k)$ can be
easily determined from the above equation. Inserting the result into
Eq.~(\ref{quickps}), one arrives at
\begin{equation}
k^3P_{\zeta }\propto k^{5+2\beta }\vert A_1(k)\vert ^2\, .
\end{equation}
For the de Sitter case, $\beta =-2$, one obtains $k^3P_{\zeta }\propto
k\vert A_1(k)\vert ^2 \propto k^0$ because $A_1(k)\propto k^{-1/2}$,
{\it i.e.}  a scale invariant spectrum. The role of the adiabatic
initial conditions, namely the fact that $A_1(k)\propto k^{-1/2}$, is
clearly crucial in order to obtain this result.

\subsection{The slow-roll power spectra}

We now evaluate the power spectra of density perturbations and
gravitational waves at leading order in the slow-roll approximation
which gives a more accurate description than the previous
back-to-the-envelop calculation. For this purpose, the details of
region II, see Fig.~\ref{potdp}, are now taken into
account~\cite{MS2}. Instead of matching the mode function of region I
directly to the mode function of region III, we now carefully
calculate the mode function in region II and perform two
matchings. The first one is between the modes function of regions I
and II and the second one is between the mode functions of regions II
and III. In order to evaluate the mode function of region II only the
calculation of the effective potentials is necessary. Using the
definitions of the slow-roll parameters, one can show that $U_{_{\rm
S}}(\eta )$ can be re-written as
\begin{equation}
U_{_{\rm S}}(\eta )\equiv \frac{(a\sqrt{\gamma })''}{a\sqrt{\gamma }}
=a^2H^2\left[2-\epsilon +(\epsilon -\delta )(3-\delta )+\xi \right]\, .
\end{equation}
It has already been established that, at leading order, $aH=-(1+\epsilon )
\eta ^{-1}$. Therefore, one obtains that
\begin{equation}
U_{_{\rm S}}(\eta )\simeq \frac{2+6\epsilon -3\delta }{\eta ^2}\, ,
\end{equation}
where one recalls that $\epsilon $ and $\delta $ should be considered
as constant. As announced before, the potential has the form $U_{_{\rm
S}}\propto {\cal O}(1)\eta ^{-2}$. For gravitational waves, similar
considerations lead to $U_{_{\rm T}}=(2+3\epsilon )/\eta ^2$. Then,
the crucial point is that the mode function can be found exactly. It
is given in terms of Bessel functions
\begin{equation}
\mu _{_{\rm II}}(\eta )=\sqrt{k\eta }\left[B_1(k)J_{\nu }
\left(k\eta \right)+B_2J_{-\nu }\left(k\eta \right)\right]\, ,
\end{equation}
where the orders are now functions of the slow-roll parameters, $\nu
_{_{\rm S}}=-3/2-2\epsilon+\delta $ and $\nu _{_{\rm T}}=-3/2-\epsilon
$. Performing the two matchings and expanding everything at the
leading order in the slow-roll parameters, one obtains
\begin{eqnarray}
k^3P_{\zeta } &=& {H^2\over \pi \epsilon \mP^2}
\biggl[1-2\left(C+1\right)\epsilon -2C\left(\epsilon -\delta \right)
\nonumber \\
& & -2\left(2\epsilon -\delta \right)\ln {k\over k_*}\biggr]\, ,
\\
k^3P_{h} &=& {16 H^2\over \pi \mP^2}
\left[1-2\left(C+1\right)\epsilon -2\epsilon \ln {k\over k_*}\right]\, ,
\end{eqnarray}
where $C$ is a numerical constant, $C\simeq -0.73$. Several remarks
are in order at this point. Firstly, the amplitude of the scalar power
spectrum depends on the Hubble parameters during inflation and on the
first slow parameter, {\it i.e.} $H^2/(\pi \epsilon \mP^2)$, while the
amplitude of the tensor power spectrum only depends on the scale of
inflation, $16H^2/(\pi \mP^2)$. The ratio of tensor over scalar is
just given by $16\epsilon $. This means that the gravitational are
always sub-dominant and that, when we measure the CMBR anisotropies,
we essentially see the scalar modes. This is rather unfortunate
because this implies that one cannot measure the energy scale of
inflation since the amplitude of the scalar power spectrum also
depends on the slow-roll parameter $\epsilon $. Only an independent
measure of the gravitational waves contribution could allow us to
break this degeneracy. Secondly, the spectral indices are given by
\begin{equation}
\nS=1-2\epsilon _1-\epsilon _2\, ,\quad n_{_{\rm T}}=-2\epsilon _1\, .
\end{equation}
As expected, the power spectra are always close to scale invariance
and the deviation from it is controlled by the magnitude of the two
slow-roll parameters. Thirdly, at the next-to-leading order there is
no running of the spectral indices since $\alpha _{_{\rm S}}$ and
$\alpha _{_{\rm T}}$ are in fact second order in the slow-roll
parameters.

\section{Inflationary predictions}

We now calculate the slow-roll parameters for typical models 
of inflation~\cite{KM}.

\subsection{Large field models}

These models typically appear in the chaotic inflationary
scenario. The potential is simply given by a monomial of the inflaton
field
\begin{eqnarray}
V(\phi _\zero) &=& M^{4}\biggl(\frac{\phi _\zero}{\mP}\biggr)^{p}\, .
\end{eqnarray} 
The calculation of the slow-roll parameters is then straightforward if
one uses Eqs.~(\ref{srpotential}). One obtains
\begin{equation}
\epsilon ={p^2\over 16\pi }\left(\frac{\phi _\zero }{\mP}\right)^{-2}\, ,
\quad 
\delta =\frac{p(p-2)}{16\pi }\left(\frac{\phi _\zero }{\mP}\right)^{-2}\, .
\end{equation}
To go further, it is convenient to express the slow-roll parameters at
Hubble crossing (remember that, at leading order, they must be
considered as constant) or, equivalently, in terms of the number of
e-folds $N_*$ between the Hubble radius exit and the end of inflation
(not to be confused with the total number of e-folds). The number of
e-folds $N_*$ is given by the formula
\begin{equation}
\label{efold}
N_*=\ln \biggl(\frac{a_{\rm end}}{a_*}\biggr)
\simeq 
-\frac{8\pi }{m_{_{\rm Pl}}^2}\int _{\phi _*}
^{\phi _{\rm end}} {\rm d}\phi _\zero V(\phi _\zero) 
\biggl(\frac{{\rm d}V}{{\rm d}\phi_\zero}\biggr)^{-1}\, ,
\end{equation}
where $\phi _{\rm end}$ is the value of the field at the end of
inflation and $\phi _*$ the value of the field at Hubble radius
crossing. Inflation stops when $\epsilon =1$ which, for chaotic
inflation, is equivalent to $\varphi _{\rm end}=m_{_{\rm
Pl}}p/(4\sqrt{\pi })$. The above integral can easily be performed and,
using the explicit expression of $\phi _{\rm end}$, one arrives at
\begin{equation}
N_*=-\frac{8\pi }{m_{_{\rm Pl}}^2}\frac{1}{p}
\int _{\phi _*}^{\phi _{\rm end}}{\rm d}\phi _\zero \phi _\zero \, \quad
\Rightarrow 
\frac{\phi _*^2}{\mP ^2}=\frac{p}{4\pi }
\biggl(N_*+\frac{p}{4}\biggr)\, .
\end{equation}
Inserting this formula into the equations giving the two slow-roll 
parameters, one obtains
\begin{equation}
\epsilon _1=\frac{p}{4(N_*+p/4)}, \quad 
\epsilon _2=\frac{1}{(N_*+p/4)}\, .
\end{equation}
Therefore, in the space $(\epsilon _1,\epsilon _2)$, a given model is
represented by the straight line $\epsilon _1=(p/4)\epsilon
_2$. However, in order to know precisely where a given model lies on
the straight line requires the knowledge of $N_*$ which, in turns,
depends on the parameters describing inflation like, for instance, the
energy scale of inflation or the reheating temperature. We will come
back to this point below.

\subsection{Small field models}

These models are characterized by potentials the shape of which, for
small values of the inflaton field, can be approximated by the
following equation
\begin{eqnarray}
V(\phi _\zero) &=& M^{4}\biggl[1-
\biggl(\frac{\phi _\zero}{\mu}\biggr)^{p}\biggr]\, .
\end{eqnarray} 
We assume that inflation takes place for values of the field smaller
than the characteristic scale $\mu $, {\it i.e.} $\phi _\zero \ll \mu
$. The two slow-roll parameters are given by
\begin{eqnarray}
\epsilon &=& {p^2\over 16\pi }\left({\mP \over \mu }\right)^2
{\left(\phi _\zero /\mu \right)^{2(p-1)}
\over \left[1-\left(\phi _\zero /\mu \right)^p\right]^2}\, ,
\\
\delta &=& -\epsilon -{p(p-1)\over 8\pi }\left({\mP \over \mu }\right)^2
{\left(\phi _\zero /\mu\right)^{p-2}\over 
\left[1-\left(\phi _\zero /\mu\right)^p\right]}\, .
\end{eqnarray}
The value at which inflation stops is given by $\phi _{\rm end}/\mu
\simeq \left(16\pi /p^2\right)^{1/(2p-2)}\left(\mu
/\mP\right)^{1/(p-1)}$. The next step is to express everything in
terms of $N_*$. Here, the model $p=2$ requires a special treatment 
and we start with this case. The integral giving the number of e-folds 
can be performed explicitly  
\begin{equation}
N_*=4\pi \left({\mu \over \mP}\right)^2
\int _{\phi _*/\mu }^{\phi _{\rm end}/\mu }{\rm d}x
\left(\frac{1}{x}-x\right)\, ,
\end{equation}
from which one obtains
\begin{equation} 
\frac{\phi _*}{\mu }\simeq 2\sqrt{\pi }{\mu \over \mP}
\exp \left[-{N_*\over 4\pi }
\left({\mP\over \mu }\right)^2\right]\, .
\end{equation}
From the above equation, we immediately deduce that
\begin{equation}
\epsilon _1\simeq \exp \left[-{N_*\over 2\pi }
\left({\mP\over \mu }\right)^2\right]\ll 1\, ,\quad 
\epsilon _2\simeq \frac{1}{2\pi}\left ({\mP\over \mu }\right)^2\, .
\end{equation}
We already see an important difference from the chaotic inflation
case.  Since $\epsilon _1$ is tiny, the observational properties of
the model will only be determined by the quantity $\epsilon _2$ which
is a $N_*$ independent quantity. Therefore, we do not need to
calculate $N_*$ in order to know where the model lies in the plane
$(\epsilon _1,\epsilon _2)$.

\par

Let us now turn to the cases $p>2$. The method is exactly the same, the 
only change being that the integral giving $N_*$ is now different but can 
still be performed analytically. After straightforward calculations, one 
arrives at
\begin{eqnarray}
\epsilon _1 &\simeq & {p^2\over 16\pi }\left({\mP \over \mu }\right)^2
\left[N_*{p(p-2)\over 8\pi }{\mP ^2\over \mu ^2}\right]
^{-{2(p-1)\over (p-2)}}
\ll 1\, , 
\\
\epsilon _2 &\simeq & {2\over N_*}{p-1\over p-2}\, .
\end{eqnarray}
As for the case $p=2$, the first slow-roll parameter is
negligible. However, the second slow-roll parameter is now a function
of $N_*$ as for chaotic models.

\subsection{The linear potential}

The linear potential is simply given by the expression
\begin{eqnarray}
V(\phi _\zero) &=& M^{4}\biggl[1-
\biggl(\frac{\phi _\zero}{\mu}\biggr)\biggr]\, ,
\end{eqnarray} 
and we still assume $\phi _\zero \ll \mu $. Since we have $V''=0$, we
deduce that $\delta =-\epsilon $ or $\epsilon _1=\epsilon _2/4$ which
is consistent with the result already obtained for chaotic
inflation. The first slow-roll parameter is independent of $N_*$ and
is given by $\epsilon _1=1/(16\pi )(\mP /\mu )^2$.

\subsection{The exponential potential}

This is an important potential since, in this case, everything can be
done exactly. The potential is given by
\begin{eqnarray}
V(\phi _\zero) &=& M^{4}\exp\left[{4\sqrt{\pi }\over \mP}
\sqrt{\gamma }\left(\phi _\zero-\phi _{\rm i}\right)\right]\, ,
\end{eqnarray}
The expression of the slow-roll parameters is $\epsilon =\delta
=\gamma $ which means $\epsilon _2=0$. The parameter $\gamma $ can be
written $\gamma =(2+\beta )/(1+\beta )$ where $\beta $ is the power
index of the exact scale factor $a(\eta )\propto \vert \eta \vert
^{1+\beta }$. The case $\beta =-2$ corresponds to the exact de Sitter
case for which $\epsilon =0$. In this case the amplitude of the
slow-roll density power spectrum is not valid.

\subsection{Hybrid inflation potentials}

Hybrid inflation typically proceeds with two fields, the role of the
second field being just to stop inflation. During the slow-roll phase,
the potential has the following shape
\begin{eqnarray}
V(\phi _\zero) &=& M^{4}\biggl[1+
\biggl(\frac{\phi _\zero}{\mu}\biggr)^{p}\biggr]\, ,
\end{eqnarray}
with $\phi _\zero \ll \mu $. Since another mechanism must be used in
order to stop inflation, one cannot calculate $\phi _{\rm end}$ in the
simple context considered here. However, if one assumes that $\phi
_*\ll \mu$, which is the case in concrete models of hybrid inflation,
then one can deduce some general features of the model. In particular, 
one can calculate the ratio of the two slow-roll parameters
\begin{equation}
{\epsilon _2\over \epsilon _1}=2-{4(p-1)\over p}
\left({\mu \over \phi _*}\right)^p \, .
\end{equation}
This means that this type of models are such that $\epsilon _2<0$ and
$\nS>1$.

\par

\begin{figure}[t]
\hspace*{-0.7cm}
\includegraphics*[width=9.1cm, height=8cm, angle=0]{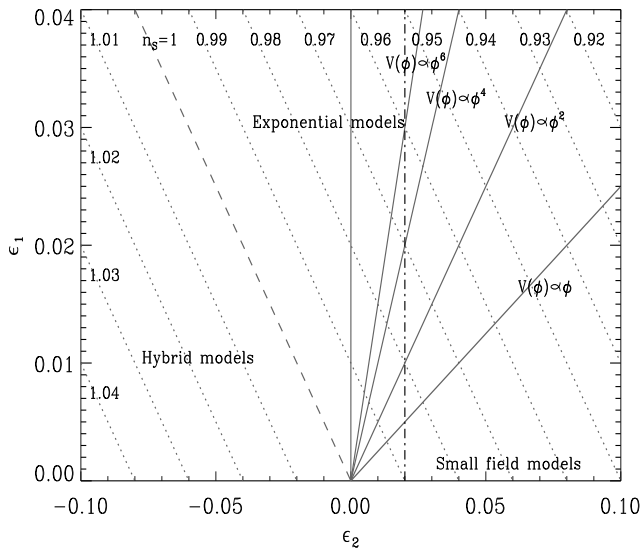}
\caption{The various models discussed in this article represented in
the plan $(\epsilon _1,\epsilon _2)$. The dotted lines are the lines
of constant spectral index. The full lines represent the location of
the large field models. The small field models are concentrated along
the $\epsilon _1=0, \epsilon _2>0$ axis whereas the exponential models
are along the $\epsilon _2=0$ line. Hybrid models have $\nS>1$ and
$\epsilon _2<0$.}
\label{srspace}
\end{figure}

The results are summarized in Fig.~\ref{srspace} where the plan
$(\epsilon _1, \epsilon _2)$ is represented.

\subsection{Comparison with the WMAP data}

\begin{figure*}[t]
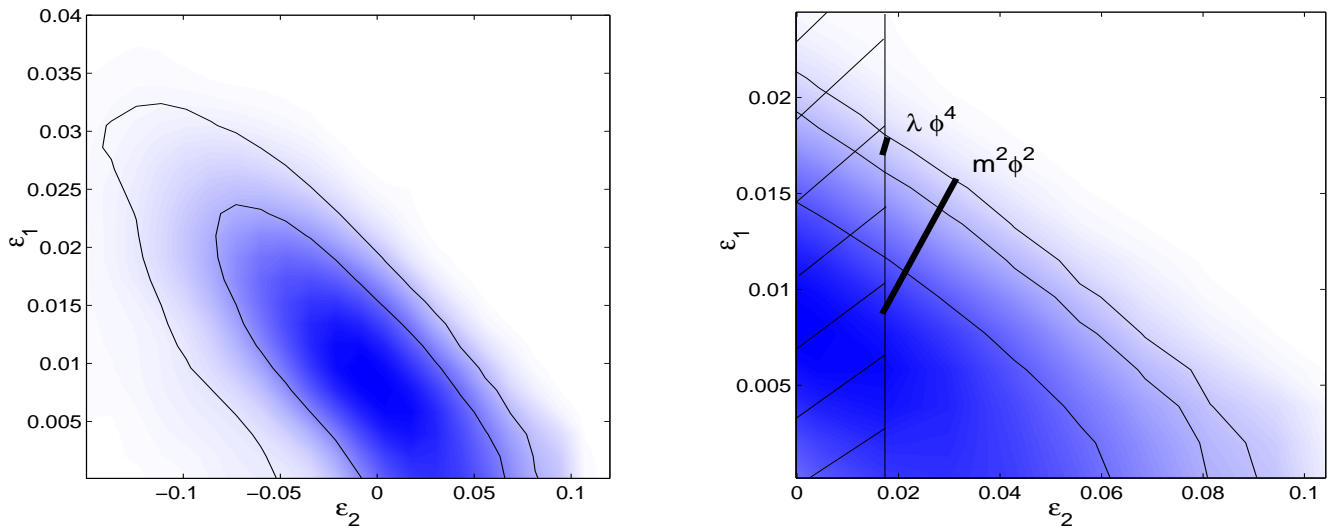

\begin{center}
\hspace*{-1.1cm}
\includegraphics*[width=8.2cm, height=7.1cm, angle=0]{fig_e1e2.epsi}
\hspace*{1cm}
\includegraphics*[width=8.3cm, height=7.0cm, angle=0]{fig_e1e22.epsi}
\caption{Left panel: allowed region in the plan $(\epsilon _1,\epsilon
_2)$ coming from the recently released WMAP data as analyzed in
Ref.~\cite{LL}. Right panel: zoom of the left panel. The model $V(\phi
_\zero ) \propto \phi _\zero ^4$ is now under big observational
pressure. These two figures are from Ref.~\cite{LL}.}
\label{e1e2}
\end{center}
\end{figure*}

The aim of this section is to illustrate the fact that the
high-accuracy data that are now at our disposal are now starting to
discriminate among the various models of inflation presented in the
last sections. The constraints coming from the most recently released
CMB data, {\it i.e.} the WMAP data~\cite{wmap}, are represented in
Fig.~\ref{e1e2} following the analysis performed in
Ref.~\cite{LL}. Very roughly speaking, we have the constraints
$\epsilon _1\leq 0.05$ and $\vert \epsilon _2\vert \leq 0.1$.

\par

As is clear from the previous analysis, except for a few models (like,
for instance, the quadratic small field model or the exponential
potential), the determination of the slow-roll parameters requires the
calculation of $N_*$ which in turns demands the knowledge of the whole
history of the universe. Unfortunately all the details of this history
are not known and hence there exits important uncertainties with
regards to the precise value of $N_*$ in a particular model. This
question has been recently re-analyzed in Ref.~\cite{LL2}. 

\par

Here, we just consider some examples to illustrate that, nevertheless,
there exists now stringent constraints on the models. For instance,
for the quadratic small field model, $\vert \epsilon _2\vert \alt0.1$
implies that $\mu /\mP \agt 1$ which is problematic for this
model. For the exponential model, $\epsilon _1\alt0.05$ means that
$\beta \alt -2.053$. In terms of the equation of state parameter, this
means (since one must have $\beta \leq -2$), $-1\alt p/\rho \alt
-0.966$ during inflation. More importantly, the chaotic models are now
severely constrained. In Refs.~\cite{LL,LL2}, it has been shown that
there is a limit on $N_*$ which implies that $\epsilon _2\agt 0.02$
(for chaotic models). This means that the models $V\propto \phi _\zero
^n$, with $n\geq 4$ are now under big pressure, as summarized in the
right panel in Fig.~\ref{e1e2}. Other important conclusions can be
obtained (in particular on the energy scale of inflation) and we refer
the reader to Ref.~\cite{LL} for more details.

\section{Open issues for inflation and conclusions}

Despite its impressive successes, inflation has problems. For
instance, it would be clearly desirable to embed slow-roll inflation
into a realistic model of particle physics at high energies (SUSY,
SUGRA, string theory etc \dots ). Unfortunately, no obvious candidate
has yet emerged (for a complete discussion of the model building
problem, see Ref.~\cite{LR}).

\par

Yet another open issue is the so-called trans-Planckian problem of
inflation~\cite{MB}. This is the fact that, at the beginning of
inflation, the scales of astrophysical relevance today were smaller
than the Planck length, {\it i.e.} where in a regime where quantum
field theory is expected to break down. Since the standard calculation
of the power spectrum is based on quantum field theory, this questions
the validity of its derivation. It is not easy to predict to which
modifications this could give rise since quantum gravity is not
known. Fortunately, one can show that ``reasonable'' modifications of
high-energy physics can leave an imprint on the observables~\cite{MR}
like the CMBR multipole moments. Therefore, there is a hope to
constrain the new physics with (future) high accuracy cosmological
observations.  This would be a concrete realization of the idea that
cosmology can help us to understand high energy physics.

\bigskip
\noindent{\bf Acknowledgments}
\medskip

I would like to thank P.~Peter and D.~Schwarz for useful discussions
and careful reading of the manuscript. I am especially indebted to
S.~Leach and A.~Liddle for the authorization to use the figures of
their article~\cite{LL}.

\end{document}